\documentstyle[12pt,psfig]{article}

\newcommand\cc{\hbox{\rm \bf c}}
\newcommand\rr{\hbox{\rm \bf r}}

\begin{document}

\begin{center}
{\large \bf Thermodynamic formalism and localization
in Lorentz gases and hopping models}
\vskip 1cm
{C. Appert$\hbox{}^{*}$\footnote{permanent address: CNRS, LPS,
Ecole Normale Sup\'erieure, 24 rue Lhomond,
75231 PARIS Cedex 05, France},
H. van Beijeren$\hbox{}^{*}$,
M.H. Ernst$\hbox{}^{*}$,
J.R. Dorfman$\hbox{}^{+}$\\
${(*)}$ Instituut voor Theoretische Fysica, Univ. Utrecht\\
Postbus 80 006, 3508 TA UTRECHT, The Netherlands\\
appert@fys.ruu.nl\\
${(+)}$ Dept. of Physics and IPST, University of Maryland\\
College Park, MD 20742, USA\\(\today)
}
\end{center}

\vskip 0.5 cm

\begin{abstract}
The thermodynamic formalism expresses chaotic properties of dynamical
systems in terms of the Ruelle pressure $\psi(\beta)$.
The inverse-temperature like variable $\beta$ allows one to scan the
structure of the probability distribution in the dynamic phase space.
This formalism is applied here to a Lorentz Lattice Gas, where a particle moving on a
lattice of size $L^d$ collides with fixed
scatterers placed at random locations.
Here we give rigorous arguments that the Ruelle pressure in the limit of infinite
systems
has two branches joining with a slope discontinuity
at $\beta = 1$. The low
and high $\beta$--branches correspond to localization
of trajectories on respectively the ``most chaotic''
(highest density) region, and the ``most deterministic''
(lowest density) region, i.e.
$\psi(\beta)$ is completely controlled by rare fluctuations
in the distribution of scatterers on the lattice, and
it does not carry any information on the global structure
of the static disorder.

As $\beta$ approaches unity from either side,
a localization--delocalization transition leads to a state
where trajectories are extended and carry information
on transport properties.
At finite $L$ the narrow region around $\beta = 1$
where the trajectories are extended scales
as $(\ln L)^{-\alpha}$, where
$\alpha$ depends 
on the sign of $1-\beta$,
if $d>1$, and as $(L\ln L)^{-1}$ if $d=1$.
This result appears to be general for diffusive
systems with static disorder, such as random walks
in random environments or for the continuous Lorentz
gas.
Other models of random walks on disordered
lattices, showing the same phenomenon,
are discussed.
\end{abstract}


\vskip 2.5 cm
\section{Introduction}

In the past several years a large body of research has focused on the
problem of relating the macroscopic behavior of non-equilibrium
systems to the underlying chaotic dynamics of the particles of which
the system is composed.
Some macroscopic transport coefficients appearing in hydrodynamic-like
equations have been related to microscopic quantities which characterize
the chaotic properties of the system
\cite{gaspard-n90,dorfman-g95, gaspard-d95,evans-c-m90}.
Ruelle, Sinai, and Bowen \cite{ruelle78,beck-s} introduced
a powerful method to derive most
of the interesting chaotic properties of a given system from a
free-energy type function, called the Ruelle or topological pressure.
This {\em thermodynamic formalism} is based on
a partition function calculated in a dynamical phase space.
For systems governed by discrete, rather than continuous
dynamics,
one point in the dynamical phase space over $t$ time steps
consists of a trajectory,
$\Omega(t) = \{x_1, x_2, \cdots, x_t\}$,
which is a set of $t$ successive states of the system.
The topological pressure $\psi(\beta)$ is 
defined
as the infinite time limit of the logarithm of the partition function
divided by the time $t$, in a way similar to the definition of the
free-energy per particle in a canonical ensemble, in the
thermodynamic limit. Again, in analogy with the methods of equilibrium
statistical mechanics, there is an
inverse temperature--like parameter $\beta$ which allows one
to scan the structure of the probability distribution
for $\Omega$. 

This formalism has been successfully applied to
Lorentz gases \cite{gaspard-b95,ernst95,dorfman-e-j95,ernst-d94}. These are models in which independent
light particles are moving among fixed scatterers.
They can be considered as elementary models for 
diffusive transport in fluids and solids.
In the continuous case, the effects of disorder in the
configuration of scatterers can be taken into account and chaotic
properties can be computed in the region of $\beta \approx 1$, as will be discussed in another paper \cite{latz}.
The model can be simplified further by constraining the light
particles to move on a regular lattice, and placing scatterers,
with some density, at random locations on the sites of this lattice.
Such models are called Lorentz Lattice Gases (LLG's).
For some of these we already calculated
the Ruelle pressure around $\beta = 1$ in the framework
of a mean-field approximation
\cite{ernst95,dorfman-e-j95,ernst-d94}.
For one-dimensional open systems on a lattice of size $L$ we obtained 
the escape rate $\gamma$, 
the Lyapunov exponent $\lambda$, and the Kolmogorov-Sinai (KS) entropy,
and found good agreement with independent direct numerical estimates
of the same quantities \cite{ernst95}.
All these quantities depend on the average density of scatterers
$\rho$.

We have previously reported \cite{appert96b} that, for $\beta$ different
from unity, and for systems with static disorder, the Ruelle pressure has unexpected properties as $L$ becomes large enough.
In particular, in the thermodynamic limit, $L \to
\infty$, it becomes independent of the density
of scatterers.
The cause of this is that, for large systems,
the Ruelle pressure is completely determined by rare
localized fluctuations in the configuration of scatterers.
This peculiar behavior is expected to be general
for all diffusive systems with static disorder, in any dimension.
In this paper we develop the {\em analytical}
arguments allowing for such a claim.
The numerical counterpart will be presented in a separate paper
\cite{appert-e96}.

In the next sections we first describe the LLG model in more detail,
and then introduce the thermodynamic formalism. Moreover,
we will extend the results of \cite{appert96b} to a mixed random walk model,
 and use it throughout the paper to illustrate the generality of our results.
The main line of our calculation is to construct exact upper and lower
bounds for the Ruelle pressure
which, as we will show, coincide in the thermodynamic limit, and are
determined by rare configurations of scatterers, except in a small region
about $\beta=1$. We refer to this situation as localization of orbits
on rare fluctuations of disorder. This means that the dominant
contributions to the Ruelle pressure in the limit of large systems
originate from orbits (points in the dynamical phase space) where the
particle is
restricted to move on those rare fluctuations, {\it i.e.} for $\beta
< 1$, on the largest compact cluster of scatterers, and for $\beta >1$
on the largest hole. We need, as a side result, the distribution 
function of
 the largest cluster size over all configurations, and the crude estimate
of \cite{appert96b} will be refined.
The analysis of finite size effects shows that the thermodynamic limit
is approached extremely slowly, $\sim (\ln L)^{-\alpha}$
where $\alpha$ depends on the model and on the sign of $1-\beta$. 
For finite systems, we have estimated the $\beta$--range around
$\beta=1$ in which the Ruelle pressure is still determined by
trajectories extending over the whole system.
As $\beta$ is deviating more and more from unity, the orbits become
more and more localized on the largest cluster or in the largest hole
of the entire configuration. In one dimension, there is an
intermediate state with "weak" localization ( see section 7).

The extension to continuous Lorentz gases
is presented in section 8.

\section{ Lorentz Lattice Gases}

\label{sect_models}

A `light' particle moves ballistically in a finite simple cubic domain 
${\cal D}$ having periodic or absorbing boundaries and containing
 $V=L^d$ sites of a $d$-dimensional
cubic lattice.
The allowed states of the system $x=\{\rr,\cc_i\}$
at time $t$ ($t=0,1,2,\cdots$) are specified by the position
$\rr \in {\cal D}$ and the velocity
$\cc_i$ of the moving particle.
The set of possible velocities $\cc_i$ ($i=1,2,\cdots,b$)
connects each site to its $b$ nearest neighbors,
where the coordination number $b$ equals $2d$ for
a 
simple 
hypercubic lattice. A fraction $\rho$ of the sites
-- chosen at random -- is occupied by a scatterer
or `heavy' particle.
The quenched configuration of scatterers is specified by the
set of Boolean variables $\{\hat{\rho}(\rr),\; \rr \in {\cal D}\}$,
where $\hat{\rho}(\rr) = 1$ if site $\rr$ is occupied by a scatterer,
and $\hat{\rho}(\rr) = 0$ if site $\rr$ is empty.
When the light particle hits
a scatterer, it is scattered to one of the
lattice directions with a probability that depends on its
incident velocity.
The scattering laws are further specified by introducing
a transmission coefficient $p$, a reflection coefficient $q$,
and, for hypercubic lattices, a deflection coefficient
$s$, normalized as
\begin{equation}
p + q + 2(d-1)s = 1.
\end{equation}
More formally, $W_{ij}$ with $i,j = \{1,2,\cdots,b\}$
is the probability that the moving particle with incident
velocity $\cc_j$ is scattered to the outgoing velocity
$\cc_i$ with normalization $\sum_i W_{ij} = 1$. For instance,
on a square lattice, the transition matrix has the form
\begin{equation} \label{a2}
W_{ij} = \left(\begin{array}{cccc}
p & s & q & s \\
s & p & s & q \\
q & s & p & s \\
s & q & s & p
\end{array} \right).
\end{equation}

The scattering at site $\rr$ is described by the random
transition matrix $\hat{W}_{ij}(\rr)$ which depends on the
configuration of scatterers $\{\hat{\rho}(\rr)\, ; \;\;
\rr \in {\cal D}\}$,
and is given by
\begin{equation} \label{a3}
\hat{W}_{ij}(\rr) = \hat{\rho}(\rr)W_{ij} +
( 1-\hat{\rho}(\rr)) \delta_{ij}.
\end{equation}
At full coverage $(\rho=1)$ the moving particle performs
a random walk with correlated jumps,
referred to as the Persistent Random Walk (PRW) \cite{Haus-Kehr}.

The time evolution of this system, in a fixed configuration
of scatterers,
is described by the Chapman--Kolmogorov equation
for the probability $\pi(x,t)$, with
$x=\{\rr,\cc_i\}$, to find the moving particle at time $t$
on site $\rr$ with incident velocity $\cc_i$, i.e.
\begin{equation}
\pi(x,t+1) = \sum_y w(x|y) \pi(y,t).
\end{equation}
In the case of absorbing boundary conditions, boundary states
 $y=\{ {\bf r}^\prime, {\bf c}_j \}$ referring to a particle entering the domain
${\cal D}$ are excluded from the $y$--summation.
The transition matrix $w(x|y)$ represents the probability
to go from state $y=\{\rr^\prime, \cc_j \}$ to state
$x=\{\rr,\cc_i\}$, and is given by
\begin{equation}
w(x|y) = \delta(\rr - \cc_i, \rr^\prime ) \hat{W}_{ij}(\rr^\prime).
\label{defw}
\end{equation}

The basic ideas  of this paper are applicable to the much
wider class of diffusive models with static disorder, 
such as hopping models with bond or site disorder
\cite{acedo-e96}, as well as to continuous Lorentz gases 
(see section 8).
As the most immediate generalization of a LLG,
we consider another model of a random walk,
called a mixed random walk (MRW),
in which a particle moves on a lattice filled by a random mixture of
{\cal X} types of scatterers. This model may be described by
{\cal X} scattering matrices of the form of
Eq.(\ref{a2}), i.e. $W_{ij}^{(k)}$ with parameters
 $p_k,q_k,s_k$ ($1 \leq k \leq$ {\cal X}).
The model contains the `ballistic' LLG, described above, as the 
special case with {\cal X} = 2, $p_1 = 1-q_1 = p$, and
 $p_2 = 1-q_2 = 1$ or $W^{(2)}_{ij} = \delta_{ij}$.
The scattering at site $\rr$ in the MRW-model is then described
by the random transition matrix
\begin{equation}
\hat{W}_{ij}(\rr) = \sum_{k=1}^{\rm X}
\hat{\rho}_k(\rr) W^{(k)}_{ij}.
\label{defW_MRW}
\end{equation}
where $\hat{\rho}_k(\rr) = 1$ if site $\rr$
is occupied by a scatterer of type $k$, and zero otherwise.

Boundary conditions may be either periodic (closed system)
or absorbing (open system) on the boundaries of domain 
${\cal D}$, 
and the transition matrix satisfies the normalization relations
\begin{equation} \label{a7}
\sum_x w(x|y) \left\{\begin{array}{cc}
= 1 & \mbox{(closed)} \\
\leq 1 & \mbox{(open)}.
\end{array}
\right.
\end{equation}
The inequality sign in Eq.(\ref{a7}) for {\em open} systems refers to the
case where $y = \{ \rr,\cc_i \} $ denotes a
state at a boundary site $\rr$ with non--entering
velocity (boundary states with entering velocity do not occur).
Indeed, the sum over $x$
excludes states where the particle has escaped from 
the domain ${\cal D}$. Hence the probability for remaining inside the
domain decreases when the particle finds itself on a boundary site.

\section{Thermodynamic formalism}
\label{sect_thermoform}

As stated in the introduction, the starting point
for this paper is a partition function defined in the dynamic
phase space whose points $\Omega(t)$ represent trajectories of
$t$ time steps :
\begin{equation} 
Z_L\left( \beta ,t|x_0\right)
 = \sum_\Omega [P\left( \Omega ,t|x_0\right)] ^\beta,
\label{defZ}
\end{equation}
where $P\left( \Omega ,t|x_0\right) $ is the probability that the system 
follows a trajectory
$\Omega(t) = \{x_1, x_2, \cdots, x_t\}$,
starting from $x_0$ at $t=0$ 
in a given system of linear dimension $L = V^{1/d}$.
The temperature-like
parameter $\beta $  allows us to scan the structure of the probability 
distribution $P$, where large positive and negative $\beta$-values select 
respectively the most probable and most improbable trajectories
\cite{beck-s}. The concepts used in this section have been discussed
in great detail in Ref.\cite{gaspard-d95,gaspard-b95}.

In each specific system
the probability $P(\Omega, t | x_0)$ for a given
trajectory can be expressed in terms of the transition
probabilities $w(x|y)$:
\begin{equation}
P(\Omega, t | x_0) = \Pi_{n=1}^{t} w(x_{n}|x_{n-1}).
\label{prodP}
\end{equation}
The partition function is determined
by the properties of the matrix
$w_\beta(x|y) \equiv [w(x|y)]^\beta$, which is defined by raising each
matrix element $w(x|y)$ to the power $\beta$.
For large times the partition function for almost all systems
becomes independent of the
initial point $x_0$ (ergodicity, see \cite{ernst-d94}), and is
determined by the largest positive eigenvalue $\Lambda_L(\beta)$ of the matrix $w_\beta(x|y)$,
which for ergodic systems can be shown to be non--degenerate.

There is a slight complication because hypercubic lattices
are bipartite, i.e. the moving particle is always on
even sites at even times and on odd sites at odd times,
or vice versa.
The system therefore consists of two independent ergodic
components, which should be considered separately, 
and the matrix $w_\beta(x|y)$ is called a periodic matrix of period 
two  \cite{Feller}.
To avoid these complications one may consider the time
$t$ to be an even integer multiple of the time step
and then consider the matrix
$w^2_\beta(x|y)=\sum_{z}w_\beta(x|z)w_\beta(z|y)$,
defined between even or between odd sites only, to be
the fundamental matrix.
Again, the largest eigenvalue $[\Lambda_L(\beta)]^2$ of the matrix
$w^2_\beta(x|y)$, restricted to one sublattice, is 
non--degenerate.

In addition, the topological or Ruelle pressure
is defined as
\begin{equation}
\psi _L\left( \beta, \rho \right) =\lim_{t\rightarrow \infty }\frac 1t
\left\langle \ln Z_L\left( \beta ,t|x_0\right) \right\rangle _{\rho},
\label{a10}
\end{equation}
where $\langle \ldots \rangle_{\rho}$ denotes an average over all 
configurations generated by the prescription that for each lattice
site, independently, $\rho$ is the probability that it will be
occupied by a scatterer. The
topological pressure is independent 
of $x_0$, if the system is ergodic, and  can be expressed in terms of the 
largest eigenvalue, $\Lambda_{L}(\beta)$, of the matrix $w_\beta$ as
\begin{equation} \label{a11}
\psi_L(\beta,\rho) = \langle\ln (\Lambda_L(\beta)) \rangle_\rho,
\end{equation}
where we have taken the infinite time limit inside the configurational
average.

Several chaotic quantities can be derived from this
function \cite{gaspard-d95,gaspard-b95}.
For example: the sum of all {\em positive} Lyapunov exponents
is $\lambda \equiv \sum_i^{(+)} \lambda_i = - \psi_L^\prime (1)$;
the escape rate for open systems is
$\gamma = - \psi_L (1)$;
the Kolmogorov-Sinai entropy follows from the generalization
of Pesin's theorem to
$h_{KS} = \psi_L(1) - \psi_L^\prime (1)$;
the topological entropy $h_T$ satisfies $h_T = \psi_L (0)$;
the Hausdorff dimension $d_H$ of the repeller
(the set of trajectories that never escape)
for an open system is the zero--point of the Ruelle pressure,
i.e. $\psi_L(d_H) = 0$.
A prime in the above formulas denotes a $\beta$-derivative.

\section{Upper and Lower Bounds}
\label{sect_upper}

In this section the Ruelle pressure will be calculated 
in the limit of infinite system size,
by constructing upper and lower bounds at finite $L$ and 
analyzing their limiting behavior for large $L$.
Consider first the {\em Lorentz lattice gas}.
For a {\em closed} system it follows from the definition of 
$w_\beta (x | y)$ and Eq. (\ref{a3}) that 
\begin{equation} \label{c1}
\sum_x w_\beta (x | y) = \hat{\rho} (\rr) W(\beta ) +
(1-\hat{\rho}(\rr) )
\end{equation}
with $y = \{\rr, \cc_i \}$ and
\begin{eqnarray} \label{c2}
&W(\beta ) \equiv  a + b + 2(d-1)c &\nonumber \\
&a \equiv p^{\beta}, \quad b \equiv q^\beta ; \quad c \equiv s^\beta.&
\end{eqnarray}
For {\em open} systems the equality sign in Eq.(\ref{c1})  is replaced by a
`less than' sign in case $y$ is a boundary state.
As a general upper bound, valid for both open and closed systems, we obtain
\begin{eqnarray} \label{c3}
\sum_x w_\beta (x | y) &\leq W(\beta ) \qquad & (\rr = {\rm scattering \;\;
site}) \nonumber \\
\sum_x w_\beta (x | y) &\leq 1 \qquad & (\rr = {\rm empty \;\;
site}).
\end{eqnarray}
If $\beta < 1$, this implies that
$\sum_x w_\beta(x|y) \leq W(\beta )$ everywhere,
as $W(\beta ) \geq 1$. This inequality combined
with Eqs.(\ref{defZ}) and (\ref{prodP}) yields:
\begin{equation} \label{c4}
Z_L(\beta,t|x_0) \;\;\leq \;\;W(\beta )
Z_L(\beta , t-1 | x_0 ) \;\;\leq \;\;(W(\beta ))^t.
\end{equation}
Then the pressure, defined by Eq.(\ref{a10}),
satisfies the inequality:
\begin{equation}
\psi_L(\beta,\rho) \leq \ln W(\beta).
\label{up1}
\end{equation}
If  $\beta>1$, and consequently $W(\beta) \le 1$, 
the analog of Eq.(\ref{c4}) becomes :
\begin{equation}
Z_L(\beta,t|x_0) \leq 1
\end{equation}
and 
\begin{equation}
\psi_L(\beta,\rho) \leq 0.
\label{up2}
\end{equation}
The above equations, (16) and (18), provide  upper bounds
for the Ruelle pressure for all $\beta$--values.

To construct a lower bound to $Z_L$ we consider clusters of scatterers,
 i.e. regions where every site is occupied by a scatterer,
and select the cluster of largest size $M$.
A cluster is said to have {\it size}
$M$, if the largest 
inscribed cube, oriented along 
the lattice directions, has a linear dimension $M$.
If there are several largest 
clusters of the same size, just choose one arbitrarily. The value of $M$
is well defined for any given configuration of scatterers.

As all terms in the sum (\ref{defZ}) are non--negative,
any sum over a subset of trajectories will give
an exact lower bound for $Z_L$.
Let $Z_M^{{\rm RW}}(\beta,t|x_0)$
denote the sum over all trajectories
which remain confined for $t$ time steps
to the largest 
inscribed cube of size $M$
(again, if for a given cluster there is more than one
inscribed cube of linear size $M$, choose one arbitrarily),
then we have a lower bound: 
\begin{equation} 
Z_L \geq Z_M^{{\rm RW}}.
\label{lowerZRW}
\end{equation}
In fact, $Z_M^{{\rm RW}}$ is equal to the partition function of a PRW in an 
open hypercubic domain with $M^d$ sites.
According to Eq.(\ref{a11}) this requires the largest eigenvalue
$\Lambda^{\rm RW}_{M} (\beta )$ of the matrix $w_\beta (x | y )$ for
the PRW, which can be found in Ref.\cite {e-d96}, and reads for sufficiently large
$M$-values :
\begin{equation} \label{c5}
\Lambda^{\rm RW}_{M} (\beta ) = W(\beta ) \{ 1 - \Delta (\beta ) k^2 + {\cal{O}}
(k^3) \}
\end{equation}
where $k^2 = \sum_{\alpha = 1}^d k_\alpha^2$ and
\begin{equation} \label{c6}
\Delta (\beta ) = \left(\frac{1}{2d}\right) \frac{a+(d-1)c}{b+(d-1)c}.
\end{equation}
Here $k$ is the smallest wave number accessible to the system,
i.e. $k=0$ for a closed system (with periodic boundaries) and 
$k_\alpha \simeq \pi/M$ ($\alpha = x,y,\cdots,d$)
for an open hypercube (with absorbing boundaries). 
On the basis of Eqs.~(\ref{a11}) and (\ref{lowerZRW}) we find the first lower 
bound on the Ruelle pressure, valid for all $\beta$--values:
\begin{eqnarray} \label{c7}
\psi_L (\beta , \rho ) & \geq & 
\langle \ln \Lambda^{\rm RW}_{M} (\beta ) \rangle_{\rho} \nonumber \\
& & = \;\; \ln W(\beta ) - \Delta (\beta ) \langle d \pi^2 / M^2 \rangle_\rho.
\end{eqnarray}
If we can show that the moment $<1/M^2>$ tends to zero when the system
size increases, then this lower bound
will tend to the upper bound (\ref{up1}) in the range $\beta < 1$. In the range 
$\beta >1$, this lower bound will approach the finite negative value 
$ \ln W(\beta)$.

We need another lower bound which will tend to the
upper bound (\ref{up2}) for $\beta > 1$.
In order to find it,
we consider for any fixed configuration
the longest line segment free of scatterers. 
Contrary to the largest cluster defined
 above, the largest empty line segment is always a
one-dimensional domain, whatever the dimensionality
of the system is.
Let $\overline{M}$ be the number of empty sites on this line segment,
 and $\overline{Z}$ the partition
sum (\ref{defZ}) restricted to trajectories confined to this line segment.
In fact, we keep only a single
trajectory, which travels 
continually through the empty region and
is reflected by the two scatterers at the end sites.
For sufficiently large times the number of reflections is
approximately $t/\overline{M}$.

The resulting sum is $\overline{Z} \simeq q^{\beta t/\overline{M}}$
and we have thus a second lower bound for the Ruelle
pressure, valid for all $\beta$--values,
\begin{equation}  \label{c8}
\psi_L(\beta,\rho) \geq
\beta (\ln q)\; \langle 1/\overline{M} \rangle_{\rho}.
\end{equation}
In summary, the following upper and lower bounds apply to all LLG's :
\begin{eqnarray} \label{c8c}
\ln W(\beta ) - \Delta (\beta ) \langle d \pi^2 / M^2
\rangle_\rho \;\leq\; \psi_L (\beta , 
\rho ) &\leq \ln W(\beta ) \qquad & (\beta < 1) \nonumber \\
\beta (\ln q) \;\langle 1/ \overline{M} \rangle_\rho 
\;\leq \; \psi_L (\beta , \rho ) &\leq \;\; 0
\qquad & ( \beta > 1)
\end{eqnarray}
The above bounds can be extended straightforwardly to the 
mixed random walk (MRW) models, where
Eq.~(\ref{c1}) becomes
\begin{equation} \label{c1bis}
\sum_x w_\beta (x | y) = \sum_{k=1}^{\rm X}
\hat{\rho}_k(\rr) W^{(k)}(\beta).
\end{equation}
with $W^{(k)}(\beta), a_k, b_k$, and $c_k$
defined in a similar way as in Eq.(\ref{c2}).

The lower bounds $Z^{\rm RW}_M$ and $\overline{Z}^{\rm RW}_{\overline{M}}$ are
respectively  determined
by the largest cube of $W^+$-scatterers containing
$M^d$ scatterers, where $W^+$ is the type of scatterers
for which $\ln W^{(k)}(\beta)$ is largest for a given $\beta$.
The resulting upper and lower bounds in MRW-models can then
be summarized as:
\begin{equation} \label{c9}
\ln W^+(\beta ) - \Delta^+(\beta ) \langle d \pi^2 / M^{2} \rangle_\rho 
\leq \psi_L (\beta , 
\rho )\; \leq\; \ln W^+(\beta ).
\end{equation}
The bounds for $\beta < 1$ contain the LLG as a special case;
the bounds for $\beta > 1$ are different.

\section{Thermodynamic Limit}

\label{sect_thermolimit}

The goal of this section is to show that the upper bounds of section 
\ref{sect_upper} are indeed the asymptotic values for
the Ruelle pressure in the limit of infinite systems ($L \to \infty$).
To do so, we need to evaluate the inverse moments, $<M^{-k}>$ $(k=1,2)$,
entering
in Eq.(\ref{c8c}), in the limit as $L \to \infty$. This requires
the asymptotic behavior of the
probability that the largest cluster is of size $M$.

We first consider the {\em one-dimensional} case where configurations
are generated by distributing scatterers on the lattice sites
according to the prescription that the occupation probability for each
lattice site is $\rho$, independently of the other sites.
Then the total number of scatterers $N$ may fluctuate
around its average value $\rho L$.
A crude estimate can be obtained
by noticing that the average number of clusters of size $m$
is approximately $L \rho^{m} (1-\rho)^2$. Indeed the cluster
can be centered on $L$ different positions on the lattice
(or $L-m$ positions for an open system),
it contains $m$ scatterers, and is bordered by two empty
sites.
For $m$ to be a typical value for the size of the largest
cluster, the above expression must be of order unity,
which implies that $M$ scales as $\ln L$.
For the inverse moments of $M$ this implies
 $<M^{-k}> \sim (\ln L)^{-k}$ for large $L$.
Hence upper and lower bounds in 
Eqs.(\ref{c8c}) approach the same limit.

This argument can be extended directly to higher dimensions. A
cluster of size $m$ (this means that the largest
inscribed cube has side $m$) occurs roughly
$L^d \rho^{m^d}$ times, where we used that for large $m$
the probability of finding at least one empty site
in each of the boundary hyperplanes is very close to unity.
For $L$ sufficiently large $L^{d}\rho^{m^{d}}$ is of order unity if $m^d \sim
\ln L$. Consequently  the inverse moments $<M^{-k}> \;\sim 
(\ln L)^{-k/d}$ for $L \to \infty$.

The $L$--dependence of the inverse moments can be obtained more
rigorously by the following observation.
We identify the clusters of size $m$, 
with $m \gg 1$, as
`noninteracting molecules' of {\em species} $m$
with partial densities \\
$\{ n(m) \simeq \rho^{m^d}; m=1,2,...\}$.
The probability to find a
volume of size $V=L^d$ unoccupied 
by clusters of size $> M$ 
is then
\begin{eqnarray} \label{d2}
P(M)& =&  \exp \left[ -
V \sum_{m>M} 
n(m) \right]  \nonumber \\
&\sim &  \exp \left[ -L^d \rho^{M^d} \right].
\end{eqnarray}
Here we 
replaced the sum in the exponential by
the first term,
since 
the size of subsequent terms in the series decreases extremely rapidly.
We 
also replaced $M+1$ by $M$, which will induce
some correction terms of relative order $( \ln L)^{-1/d}$
in the final expression (\ref{d5}).
The probability that the largest cluster is exactly of
size $M$ is then $A(M) = P(M) - P(M-1)$,
or in the continuum limit,
\begin{equation} \label{d3}
A(M) = P^{\,\prime} (M) \sim - d L^d \ln (\rho) \, M^{d-1}
\rho^{M^d} P(M).
\end{equation}
For large $L$ the inverse moments 
$<M^{-k}> = \int^\infty_0 d M M^{-k} A(M)$ can be evaluated asymptotically by a saddle point 
method, as $A(M)$ is sharply peaked around 
its maximum. The maximum is located at $M_0$, which is the root of:
\begin{equation} \label{d4}
\left[ \ln A(M) \right]^\prime =
- d L^d (\ln \rho) M^{d-1} \rho^{M^d}
+ d 
\ln (\rho) M^{d-1} + (d-1) M^{-1}.
\end{equation}
For large $M$ the solution of this equation is determined by the first two terms on the right hand side (dominant balance
argument),
yielding,
\begin{equation} \label{d5}
M_0 \simeq \left ( \frac{d \ln L}{|\ln \rho|} \right)^{1/d}
\end{equation}
with correction terms of relative order $( \ln L)^{-1/d}$.
For large $L$  the inverse moments behave asymptotically as,
\begin{equation}
<M^{-k}> \sim M_0^{-k} = \left ( \frac{d \ln L}{|\ln \rho|} \right)^{-k/d}
\label{d6}
\end{equation}
in agreement with the crude estimate above.

In the preceding paragraphs the 
probability $\rho$ of occupation of sites by scatterers has been kept fixed.
It can be shown that the results (\ref{d5}) and (\ref{d6}) for the $L$ 
dependence are still correct if one fixes the total
number of scatterers, $N$, in all configurations \cite{david-b}. 

\begin{figure}
\centerline{\psfig{file=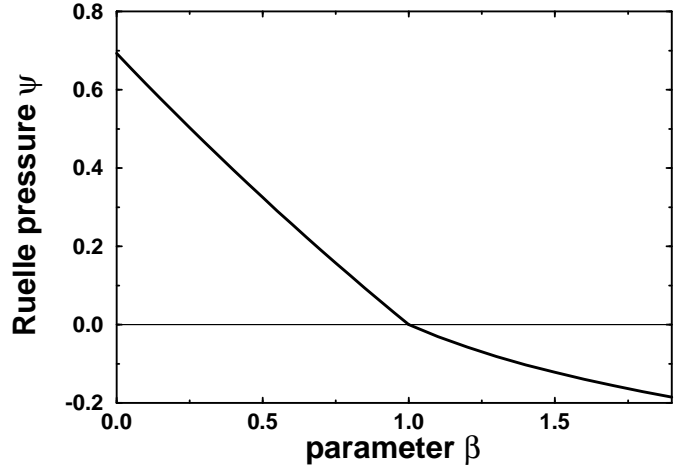,width=3.5in}}
\caption{\em Ruelle pressure for a MRW in a one-dimensional
system of
infinite size for $p_1=0.9$ and $p_2=0.7$. It is independent of the
density of scatterers as long as $\rho_1$ or $\rho_2$ is not $0$ or $1$.
Note the difference with the LLG, where the branch for $\beta>1$
is $\psi = 0$.}
\label{figpsilimit}
\end{figure}

The above results can also be used to estimate the typical size $\overline{M}$
of the largest empty line segment.
The distributions for scatterers and holes are symmetric
by exchange of $\rho$ and $1-\rho$.
As we are interested in a one-dimensional domain,
whatever the dimension $d$  of the lattice,
we have to replace $M^d$ by $\overline{M}$.
We straightforwardly obtain
\begin{equation} \label{d7}
<1/{\overline{M}}> = 1/M_0  \simeq -
\frac{\ln (1-\rho)}{d \ln L}.
\end{equation}
We conclude therefore that in the thermodynamic limit
$L \to \infty$, the moments $<M^{-k}>$ vanish and thus
the lower bounds of section 
\ref{sect_upper} converge towards the $L$--independent upper
bounds.
This yields for the Ruelle pressure in LLG's 
in the thermodynamic limit
\begin{equation} \label{d8}
\lim_{L\rightarrow \infty} \psi_L(\beta,\rho) =
\left\{ \begin{array}{ll}
\ln W(\beta )  & ( \beta < 1) \\
0	       & ( \beta > 1) .
\end{array}
\right.
\end{equation}
In the MRW-models,
the Ruelle pressure  is
\begin{equation} \label{d9}
\lim_{L\rightarrow \infty} \psi_L (\beta , \rho ) = 
\ln W^+(\beta )
\end{equation}
as is illustrated in figure \ref{figpsilimit} for MRW models.

\section{Localization, extension to MRW's}

\label{sect_localization}

In the previous section we have shown that the dynamic partition function 
and the Ruelle
pressure of LLG's in thermodynamically large systems are completely determined
 by the rare fluctuations in the spatial distribution of scatterers. It
is worth stressing  again here that for the regions $\beta <1, \beta = 1$,
and $\beta >1$, and $L \rightarrow  \infty$, different sets of
trajectories make the dominant contributions to the Ruelle
pressure. For $\beta <1$ the
trajectories contributing to the partition function in a given configuration of
scatterers are localized on the
largest compact cluster of scatterers, i.e.
localized in the `most chaotic' or `least deterministic' region. For
$\beta >1$  only a single trajectory  contributes,
which --- on any $d$--dimensional lattice --- is localized on the
longest line segment that is free of scatterers, i.e. the relevant
trajectory is localized in the `least chaotic' or `most deterministic' region
of configuration space

Therefore,
as $L \rightarrow \infty$ the Ruelle pressure 
becomes independent of the configuration
(except for atypical configurations such as strictly periodic
ones); it is even independent of the density of scatterers
(except at $\rho=0$,
where no fluctuations exist any more).
It does not carry any information 
on the structure of the random medium.

Since for $\beta$ different from unity, relevant trajectories do not
explore the whole system but only a small part of it, the `mean field
configuration' with all scatterers more or less equidistant from one
another is not at all a typical configuration. On the contrary, it is
the one that gives, among all configurations, the minimal value for
the Ruelle pressure. On the other hand, the maximal Ruelle pressure is
obtained for the configuration where all of the scatterers form a
single compact cluster. The average will be somewhere in between. This
means that any calculation starting from a mean field approximation
will give very poor results for $\beta$-values different from unity \cite{cecil-paris}.

However at $\beta=1$,
the Ruelle pressure for the LLG
and its derivatives with respect to $\beta$ do depend 
on the overall density \cite {ernst95} and on more details of the  total configuration of scatterers\cite{acedo-e96}.
For {\em finite} $L$, this is also true in a small region around $\beta = 1$.
There, relevant trajectories are extended or delocalized,
and explore large regions of the lattice.
This conclusion is based on the reasonably good agreement for escape rates and
Lyapunov exponents between the results from computer
simulations and mean field calculations for the LLG \cite{ernst95}.

The same conclusions carry over to the MRW-models, where in the thermodynamic
limit the trajectories are localized on the largest compact
cluster  with $W^+(\beta)$--scatterers.
At $\beta = 1$, all scatterers have the same $\ln W = 0$
and again trajectories are delocalized on the whole lattice.
The structure of the typical mean field configurations, 
contributing around $\beta =1$, has not yet been explored,
and mean field estimates for the Lyapunov exponents in
open systems have not yet been derived for MRW's.

Interesting new phenomena can occur near
those values of $\beta \neq 1$ where different types of scatterers may have the
same value of $\ln W$.
For example, this occurs at $\beta = 0$
if all scatterers have the same number of 
non--zero--scattering directions.
Then the moving particle cannot distinguish between the different types of
scatterers and the relevant trajectories become again ``delocalized" on a large cluster with a random
mixture of the different scatterers with the same value of $\ln W(\beta)$.
More explicitly, if there are {\cal X} types of scatterers, it may occur
that $K$ ($2 \leq K < ${\cal X}) of them have the same $\ln W^+$
strictly greater than the $\ln W$ for all other  types of scatterers.
Then the  cluster of $W^+$--scatterers on which the relevant
trajectories are localized contains a  random mixture of
these $K$ types of scatterers \cite{acedo-e96}.

Suppose now that we consider, for $d >1$, a LLG or a MRW
model that shows a percolation transition. In such a case is
important to note that the definition
for the cluster size used here is not the number of connected
sites, but the size of the largest inscribed hypercube, which 
typically is much smaller than the system size even for a percolating cluster.
Thus the percolation transitions in such models have no effect on our
considerations and the results of this paper remain valid.

\section{The delocalization region}

\label{sect_crossover}

In this section we estimate the size of the delocalization region
around $\beta = 1$ in LLG's for finite systems.
The Hausdorff dimension of the repeller (i.e. the
set of trajectories which do not escape from the system
after an infinite time) is a root of the Ruelle pressure,
\begin{equation} \label{e1}
\psi_L ( d_H ) = 0.
\end{equation}
For a large system, $d_H$ is close to unity
\cite{gaspard-b95}.
Using the facts that $ \psi_{L}(1) = -\gamma \simeq
 Dd \pi^2 / L^2 $ for a hypercubic domain in $d$--dimensions,
where $D$ is the diffusion coefficient, and
 $\psi_L^\prime (1) = -\sum_{\lambda_i>0} \lambda_i
\simeq -\lambda_\infty$ where
$\lambda_\infty$ represents the sum of positive Lyapunov
exponents in the infinite $L$--limit,
we find that the Hausdorff dimension $d_H$ for a large hypercubic domain
is in first approximation,
\begin{equation} \label{e2}
d_H \simeq 1 - \left(\frac{D}{\lambda_\infty} \right) 
\frac{d \pi^2}{L^2},
\end{equation}
where $D$ and $\lambda_\infty$ depend on the density of
 scatterers.
Therefore, as the structure of the repeller is a fundamental
feature of the system, the crossover region should extend
at least over a
$\beta$--range of order $1/L^2$.
On the other hand we have concluded that for $L
\rightarrow \infty$ and for $\beta$ different from unity,
the Ruelle pressure becomes independent of the global structure
of the disorder, as the relevant trajectories become
localized in regions of the lattice where rare
fluctuations of high or low density of scatterers occur.
This was demonstrated in previous sections
in the limit of infinite systems. Therefore as long as the mean field
value of the Ruelle pressure or of the largest eigenvalue is smaller
than the lower bound, the states of the system are localized. In fact,
numerical results \cite{appert-e96} support our
intuition that the effect of localization
can be estimated fairly well for any large but finite system
by taking the lower bounds in Eqs.
(\ref{c8c}) as estimates of the Ruelle pressure
for $\beta < 1$ or $> 1$.
Hence, for a given $L$ crossover from localized to
extended states occurs as we approach $\beta=1$ from either side, when
the mean field value equals the lower bound.

To obtain an  estimate of these crossover values,
we compare the Ruelle pressure
of the mean field configuration with the estimate for the
Ruelle pressure based on the lower bounds (see Eq.~(\ref{c7})).
It is equivalent to compare the eigenvalues of the matrix $w_\beta$
associated with a localized and with a delocalized eigenstate.

The second one is obtained for the `mean--field' configuration
in LLG's from the PRW expression (\ref{c5})
by a rescaling argument \cite{ernst95,appert-e96}.
It reads
\begin{equation} \label{e3}
\Lambda^{{\rm MF}}_L (\beta) = (W(\beta))^\rho 
\left(1-\rho\Delta(\beta) d (\pi/\rho L)^2 \right)
 + {\cal O} ({1}/{L^3})
\end{equation}
with $W(\beta)$ and $\Delta$ defined in  Eqs.(\ref{c2}) and (\ref{c6}).

For $\beta < 1$ localization takes certainly place
if the lower bound on the
Ruelle pressure is larger than the mean field value, {\it i.e.},
\begin{equation} \label{e4}
\ln W(\beta) - \Delta (\beta)d \pi^2/M^2_0 \;\;
> \;\;\ln \left( W(\beta)\right)^\rho + {\cal O}({1}/{L^2}),
\label{loc_beta_small}
\end{equation}
where the left hand side is the lower bound given in Eq. (\ref{c7})
with $<M^{-2}> \simeq M_0^{-2}$ on account
of (\ref{d6}).
 The right hand side is the mean field 
value given by Eq. (\ref{e3}).
By expanding both sides in powers of
 $\epsilon \equiv 1-\beta$, we find that the Ruelle pressure is
determined by localized trajectories only if
\begin{equation}
\epsilon > \epsilon_{-} \equiv \frac{d \pi^2 \Delta(1)}{\delta (1-\rho) M_0^2}
\simeq  \frac{ d\pi^2 \Delta(1)}{\delta (1-\rho)}
\left[ \frac{|\ln\rho|}{d \ln L}\right]^{2/d},
\label{ineq1}
\end{equation}
where $\delta = |p \ln p + q \ln q +2(d-1)s \ln s|>0$.
We note that $d_H=1-\epsilon$ in Eq.(\ref{e2}) is indeed within the
delocalized region,as was to be expected.

However, the crossover between localized and delocalized states
may involve some intermediate states. In principle one might have a ``weak
localization'' in a region of size $M_1$ (with $\ln L \ll M_1 \ll L$),
where the local density, $\rho+\Delta \rho$, is slightly larger
than  $\rho$, but where $\Delta \rho$ is large enough so that trajectories,
remaining confined to this region, dominate the Ruelle pressure.
To estimate the largest density fluctuation to be found in a region of size $M_1$ we first note that the probability for a density fluctuation $\Delta \rho$
in such a region can be estimated as $\exp (\mu \, \Delta \rho \, M_1^d)$, with $\mu$
the chemical potential of the scatterers (considered as lattice gas particles). Since the number of different regions
of this size is on the order of $L^d$ for $M_1$ in the above range, the largest density fluctuation occurring
in one of these regions follows from the requirement $L^d \exp (\mu \, \Delta 
\rho \, M_1^d) \approx 1$, or 
\begin{equation} 
-\mu\Delta \rho M_1^d \sim \ln L^d.
\label{deltarho}
\end{equation}
In a manner similar to (\ref{loc_beta_small})
we compare the mean field value $\psi^{\rm MF}_{L} \simeq \ln \left(W(\beta) \right)^\rho$ with the mean field   estimate of the Ruelle pressure corresponding to trajectories
confined in a region of average density $\rho +\Delta \rho$,
\begin{equation}
\psi^{{\rm MF}}_{M_1} \simeq
\ln \left\{\left[ W(\beta) \right]^{\rho + \Delta \rho}
\left[ 1-\Delta(\beta) \frac{ d \pi^2}{(\rho + \Delta \rho)
M_1^2} \right] \right\}.
\nonumber
\end{equation}
Expressing $\Delta \rho$ in terms of $M_1$ according to (\ref{deltarho}) and 
expanding the difference $\psi^{\rm {MF}}_{M_{1}}-\psi^{\rm MF}_L$ to
lowest order in $\epsilon$ yields a condition for ``weak localization"
in the most  dense region of size $M_1$, namely,
\begin{equation}
\psi^{{\rm MF}}_{M_1}-\psi^{MF}_{L} \simeq 
\epsilon \frac{\delta d \ln L}{M_1^d} - \frac{d \Delta(1) \pi^2}{\rho M_1^2} > 0.
\label{ruellediff}
\end{equation}
By taking  the estimate of the delocalization region
in  (\ref{ineq1}) one immediately sees that for $\epsilon \ll (\ln L)^{-2/d}$, this inequality
can only be satisfied in $d=1$. 
Localization will occur on a region of size $M_{1}$ maximizing
the difference (\ref{ruellediff}) in  Ruelle pressures.
Differentiating (\ref{ruellediff}) with respect to $M_1$
gives an $M_1$ that is proportional to $1/(\epsilon
\ln L)$. As long as this is $\ll L$, weak localization will occur. As soon as 
$\epsilon \sim  \epsilon_w \equiv 2\Delta \pi^2/(\rho \delta \,L \ln
L)$ the confinement region of the dominant trajectories  becomes
comparable to the full system and  weak localization is no longer a
meaningful concept. Hence, in
one dimension one can distinguish in addition the weak localization
regime $\epsilon_w < \epsilon < \epsilon_-$. The region of full delocalization  is narrowed down to
$1-\beta < \epsilon_w$.

For $\beta > 1$ and $d>1$, the crossover value can be determined
 by comparing the mean field estimate with
the lower bound (\ref{c8}) for the Ruelle pressure in the LLG, 
combined with Eq.~(\ref{d6}), i.e.
 \begin{equation} \label{e6}
{\beta \ln q /M_0}
> \ln (W(\beta))^\rho + {\cal O}(1/{L^2}).
\end{equation}
By expanding in powers of $\beta-1 \equiv \epsilon' $ 
we find localization for
\begin{equation}
\epsilon' > \epsilon'_+ \equiv \left(\frac{|\ln q|}{ \rho \delta} \right)
\frac {1}{M_0}
\simeq \left( \frac{|\ln q|}{d \rho \delta} \right)
\frac{|\ln(1-\rho)|}{\ln L}.
\end{equation}
For $d=1$ one can show again, by using arguments similar to those above,
that there is a region of weak localization where the density is
slightly lower than average. This region occurs for $\beta$ values given by
$C'/(L \ln L) < \beta - 1 < \epsilon'_+$, where $C'$ is a positive constant.

In summary, the Ruelle pressure in LLG models is determined by
extended or delocalized states if $\beta$ is in the interval $\{1- a_-/ (\ln L)^{2/d},$ 
$1+ a_+/(\ln L)^{1/d} \}$ , for $d > 1$, and
in the interval $\{ 1-C/(L\ln L), 1+C'/(L\ln L)\}$ for $d = 1$, where
$a_{\pm},  C$, and $C'$ are some positive constants.

\section{Extension to continuous systems}
\label{sect_continuous}

Our considerations can be extended to the case of
continuous Lorentz gases with static disorder. For $\beta
< 1$, we expect the moving particle to be localized in a region of
space with a high density of scatterers, while for $\beta > 1$
it should be localized in a large region where the density of
scatterers is zero. To provide some qualitative explanations of this
observation,
we consider a Lorentz gas with hard
spherical scatterers, the so--called non overlapping
Lorentz gas. Extension to overlapping scatterers
or  soft scatterers is possible but will not be considered
here. In the
non-overlapping Lorentz gas and $\beta <1$ the Ruelle pressure
 approaches in the
thermodynamic limit that of a closely packed system
of hard spheres of diameter $a$.
To understand this it suffices to bound the dynamical
partition function by the contribution of all trajectories
confined to a hypercubic volume of size $M^d$,
containing the centers of ${\cal N}\equiv \rho M^{d}$ scatterers.
The probability of finding such a volume in the system
can be estimated conservatively to be at least
proportional to
\begin{equation}
\frac{V}{a^d} \;
\frac{Q(N-{\cal N}, V-M^d)}{Q(N, V)}
Q({\cal N}, M^d)
\label{probavoid}
\end{equation}
By expanding the logarithm of the ratio of configurational
partition
functions in powers of the volume $M^d$ of the hypercube
(using ${\cal N} = \rho M^d$) --- as the hypercube is
a small subsystem of the total system with $(N, V)$ ---
and noting on the other hand that
the partition function $Q({\cal N}, M^d)$
of the hypercube increases exponentially in $M^d$,
we conclude that the probability in (\ref{probavoid})
is proportional to $(V/a^d) \exp (-\alpha M^d)$
with $\alpha$ some constant.

For any average density below the close-packing density
this can be made of order unity by choosing
$M^d$ proportional to $\ln (V/a^d)$,
which implies that for increasing $V$ arbitrarily
large volumes with a density arbitrarily close
to the close-packing density can be found.

It is not clear what happens to the Ruelle pressure when all of the
particles can move, as in fluids, for example, although it seems
obvious that for $\beta >1$, trajectories where the particles rarely
collide will dominate the Ruelle pressure. 

\section{Conclusion}

We conclude this paper with a number of remarks:\\

1) In this paper we have discussed the Ruelle pressure for diffusive 
models with static disorder. Our results indicate that for large 
systems and for all but a small range of
values of the inverse temperature-like parameter $\beta$, the Ruelle pressure
is determined by rare fluctuations in the configuration of scatterers,
 and consequently carries no physical information on
the chaotic scattering of the moving particle during 
its motion through the frozen--in disorder.
Only in a narrow region around $\beta = 1$
does the thermodynamic formalism yield physically relevant
information on the chaotic scattering in diffusive systems
with static disorder.
 The extension to continuous systems outlined in \cite{appert96b}
has been made explicit here without having to use the usual tools
of kinetic theory. The localization phenomena in the Ruelle pressure are in fact
most similar to the asymptotic behavior (stretched exponential
decay) of the survival probability of a random walk in
a random array of absorbing traps. The survival probability
is solely determined by the extremely rare fluctuation
that the random walk finds itself in the largest region
free of traps \cite{Donsker-V}.

In different areas of statistical physics analogous
phenomena occur, where the large time or the small
frequency/energy asymptotics are controlled by
extremely rare spatial fluctuations, such as
in Lifshitz tails \cite{Lifshitz tails}, Griffith's
singularities \cite{Griffith sing} and directed polymers
\cite{Halpin-Healy}.

2) It has been shown numerically in one dimension
\cite{appert-e96}
that for finite systems and outside the crossover region
the lower bound found for the Ruelle pressure is also
a good estimate for the pressure itself, indicating that localization
on the largest cluster indeed occurs.
We conjecture that in higher dimensions
it is the largest {\em convex} cluster inscribed
in a set of connected scattering sites that will determine
the Ruelle pressure. 

3) To allow the dynamical partition function to scan the full structure of
our diffusive models with static disorder, time should
be sufficiently large that the moving particle can explore the 
entire volume of the 
system.
 Consequently $ t >> L^2$, which determines the physically relevant 
order of 
limits.
In determining the Ruelle pressure in Eq.(\ref{defZ}) one 
 takes first the
limit $t \to \infty$ for fixed $L$, and next 
allows $L$ to tend to
$\infty$. Therefore, for a fixed system size $L$,
the trajectory has an infinite time to explore the
system and to find the largest cluster, where it will then
stay localized with a high probability. An interesting open problem
remains to explore both the time and size dependence of the dynamical
partition function, Eq. (8), to see how the various features discussed here are approached
in the limit of infinite time, but finite size; to study
diffusive behavior when the time is kept finite and the size of the
lattice is allowed to become infinite; and to study the behavior of
the dynamical partition function when both $L$ and $t$ approach infinity in some coupled manner.

Acknowledgements:
We would like to thank T. Gilbert, U. Ebert, and B. Derrida
for stimulating discussions.
One of us (C.A.) acknowledges support of the foundation
``Fundamenteel Onderzoek der Materie'' (FOM), which is
financially supported by the Dutch National Science
Foundation (NWO), and of the French  ``Centre National de la
Recherche Scientifique'' (CNRS).
J.R.D. thanks the National Science Foundation for Support
under Grant NSF-PHY-93-21312.

\bibliographystyle{unsrt}

\end{document}